\documentclass[genTeX]{nrc1}
\newcommand\beq{\begin{equation}}
\newcommand\eeq{\end{equation}}
\newcommand\bea{\begin{eqnarray}}
\newcommand\eea{\end{eqnarray}}
\newcommand\beano{\begin{eqnarray*}}
\newcommand\eeano{\end{eqnarray*}}

\newcommand\eqref[1]{(\ref{#1})}
\def\factorial{\mathchar"5021\mathopen{}\mathinner{}} 
\begin{document}

\title{Problems with Complex Actions}

\bigskip\bigskip
\author{Garnik Alexanian}
\address{Troika Dialog, Romanov Pereulok 4, Moscow 125009, Russia}
\author{R.~MacKenzie, M. B. Paranjape}
\address{Groupe de physique des particules, Universit\'e de Montr\'eal,
C.P. 6128, succ. centre-ville, Montr\'eal, QC, Canada, H3C 3J7}
\correspond{paranj@lps.umontreal.ca}
\author{Jonathan Ruel}
\address{Dept. of Physics, Harvard University, Cambridge, Massachusetts, USA, 02138}
\shortauthor{G. Alexanian, R.~MacKenzie, M. B. Paranjape, J. Ruel}
\maketitle

\begin{abstract}
We consider Euclidean functional integrals involving actions which are not exclusively real.  This situation arises, for example, when there are $t$-odd terms in the the Minkowski action.   Writing the action in terms of only real fields (which is always possible), such terms appear as explicitly imaginary terms in the Euclidean action.  The usual quanization procedure which involves finding the critical points of the action and then quantizing the spectrum of fluctuations about these critical points fails.   In the case of complex actions, there do not exist, in general, any critical points of the action on the space of real fields, the critical points are in general complex.  The proper definition of the function integral then requires the analytic continuation of the functional integration into the space of complex fields so as to pass through the complex critical points according to the method of steepest descent.  We show a simple example where this procedure can be carried out explicitly.  The procedure of finding the critical points of the real part of the action and quantizing the corresponding fluctuations, treating the (exponential of the) complex part of the action as a bounded integrable function is shown to fail in our explicit example, at least perturbatively.
\end{abstract}

\begin{resume}
Nous consid\'erons les int\'egrales fonctionnelles euclidiennes qui contiennent une action qui n'est pas compl\`etement r\'eelle. Cette situation appara\^it, par exemple, quand il existe des termes dans l’action minkowskienne qui sont impairs sous le reversement du temps $t$ . En \'ecrivant l’action uniquement en termes de champs r\'eels (ce qui est toujours possible), ces termes apparaissent de fa\c con explicitement imaginaire dans l’action euclidienne. Le processus de quantification, qui correspond d’abord \`a la recherche des points critiques de l’action puis \`a la quantification des fluctuations autour de ces points critiques, ne fonctionne pas, car il n’existe pas en g\'en\'eral de point critique d’une action complexe dans l’espace des champs r\'eels. Les points critiques se trouvent dans l’espace du prolongement analytique complexe des champs r\'eels. La d\'efinition de l’int\'egrale fonctionnelle demande alors un prolongement analytique dans l’espace des champs complexes tel que le contour d’int\'egration passe par les points critiques complexes, selon le chemin impliqu\'e par la m\'ethode du col. Nous montrons un exemple simple o\`u cette proc\'edure peut \^etre effectu\'ee explicitement. La m\'ethode alternative de consid\'erer seulement la partie r\'eelle de l’action et de quantifier les fluctuations correspondantes, traitant (l’exponentielle de) la partie imaginaire comme une fonction born\'ee int\'egrable, \'echoue dans notre exemple, du moins de fa\c con perturbative.

\end{resume}

\section{Introduction\label{sec:intro}}
Quantum field theories are quite often understood as defined by the Feynman path (functional) integral\cite{f}.   However, the usual Minkowski space path integral is not actually mathematically well defined  
\beq
{\cal I}=\int{\cal D}\phi (t)\, e^{iS\left[\phi (t)\right]/\hbar}
\eeq
where $S$ is the action functional which is real.    The integrand is unimodular while the integration domain is (usually) infinite.   Therefore the Feynman path integral cannot be absolutely convergent.  For a fuller explanation see the book by Glimm and Jaffe \cite{gj}.
We consider the action functional to be written exclusively in terms of real fields, as any complex field can be re-expressed in terms of  real fields for its real and imaginary parts.  For this paper, we restrict ourselves to bosonic fields, however we are certain that everything can be quite easily extended to include fermionic fields as well.  Actually, in one or two spatial dimensions, fermions can be taken into account in terms of purely bosonic fields, via bosonisation \cite{c}, in 1+1 dimensions or a purely statistical gauge field and a Chern-Simons term in 2+1 dimensions\cite{djt}, thus fermions are fully accounted for in these dimensions. 

In fact the actual definition of the Minkowski time Feynman path integral is obtained via analytic continuation.    One starts with a presumably well defined functional integral by initially analytically continuing the action to Euclidean time and considering (real) fields that are functions of Euclidean time:
\beq
t\rightarrow -i\tau ,\quad \phi(t)\rightarrow\phi(\tau )\\
{\cal I}\rightarrow\int{\cal D}\phi (\tau )\, e^{-S_E\left[\phi(\tau )\right]/\hbar}.
\eeq
One obtains the Minkowsi space path integral via analytically continuing back to real time.  This procedure works reasonably well.  Usually the Euclidean action is a real and positive definite functional of the Euclidean time field configuration.  However for certain cases the Euclidean action is not completely real.  Such a situation is not necessarily a fatal impediment to the definition of the functional integration.  If the real part of the Euclidean action can be used to define a measure on the space of field configurations, then the fluctuating part can be seen as simply a bounded functional that can be integrated against this measure.  However in practice this is not very useful.  

In practice, the functional integration is evaluated in perturbation theory.  The Gaussian fluctuations about the critical points of the action are quantized without approximation using the Euclidean functional integral, and the non-linear terms are treated as perturbations.  It is this procedure that fails when the Euclidean action is not real.  The critical points of a complex action are in general complex.  That is, the action is critical for values of the fields that are complex.  Thus the critical points are not attained for any real field configurations, and the quantization about these critical points cannot be addressed in terms of the strictly real fields that we have started with. 

There are two procedures that are followed in the case of complex actions.  The first procedure is to consider only the real part of the action and find its critical points.  These are in general obtained for real field configurations.  Then the fluctuations about such abridged critical points can be quantized treating the imaginary part of the action and the non-linear terms perturbatively.  The second procedure is to analytically continue the ``contour" of functional integration into the space of complex field configurations, so that it passes through the complex critical point.  Then the fluctuations of these complex field configurations about the complex critical point are quantized treating the non-linear terms perturbatively.  For some examples of cases with complex critical points see the articles \cite{ahps}, \cite{tsh}, \cite{oss}.

In this paper we show, by explicit example, that  the first procedure is wrong.  It may work in some special cases, but in general it is false.  We show that the second procedure does in fact give the correct expansion.  

\section{Complex actions\label{2}}
Complex Euclidean actions can arise for many reasons.  In general they occur due to a $t$-odd term in the Minkowski action.  Such a term, under analytic continuation to Euclidean time, will necessarily be imaginary.   Consider the action
\beq
S\left[\phi (t)\right]= S_{even}\left[\phi (t)\right]+S_{odd}\left[\phi (t)\right]
\eeq
where $S_{even}\left[\phi (t)\right]\rightarrow S_{even}\left[\phi (t)\right]$ but $S_{odd}\left[\phi (t)\right]\rightarrow -S_{odd}\left[\phi (t)\right]$ for $\phi (t)\rightarrow \phi (-t)$.  For the following examples for $S_{odd}\left[\phi (t)\right]$
\beq
S_{odd}\left[\phi (t)\right]\sim\cases{\int dt d^2x\,\epsilon^{\mu\nu\lambda}(A_\mu\partial_\nu A_\lambda +\cdots)\cr \int_{half\,S^5} dtd^4x\,\epsilon^{\mu\nu\lambda\sigma\tau}Tr (U^\dagger\partial_\mu U\cdots U^\dagger\partial_\tau U )\cr \int dt d^3x\,\epsilon^{\mu\nu\lambda\sigma}Tr( F_{\mu\nu}F_{\lambda\sigma})}
\eeq
which correspond to a Chern-Simons term in 2+1 dimensions\cite{djt}, a Wess-Zumino-Novikov-Witten terms in 3+1 dimensions\cite{wz},\cite{n},\cite{w}, and the instanton number in 3+1 dimensions respectively give rise to imaginary terms in the Euclidean action.  Indeed all of these terms contain only one temporal derivative:
\beq
\int dt\int d^dx \cdots\partial_t\cdots
\eeq
where $d$ is the spatial dimension.  On analytic continuation to Euclidean time, $t\rightarrow -i\tau$ evidently the odd part of the action is invariant, 
\beq
\int dt\int d^dx \cdots \partial_t\cdots\rightarrow\int d\tau\int d^dx \cdots\partial_\tau\cdots
\eeq
while the even part of the action satisfies
\beq
\int dt\int d^dx  L_{even}\left[\phi (t)\right]\rightarrow i\int d\tau\int d^dx  L_{even}\left[\phi (\tau)\right]
\eeq
where $L_{even}\left[\phi (\tau)\right]$ is by hypothesis a real, positive functional, which means that an overall minus sign has been factored out after the analytic continuation.  Then the Euclidean functional integral becomes
\beq
{\cal I}=\int{\cal D}\phi (\tau )\, e^{-(\int d\tau\int d^dx L_{even}\left[\phi(\tau )\right]-i\int d^dx L_{odd}\left[\phi(\tau )\right])/\hbar}.
\eeq
This Euclidean functional integral is in principle well defined, the real part of the action serves to define a measure on the space of field configurations.  However the perturbative quantization is not straightforward as the critical points of the action are in general achieved at complex values of the field configurations.  In the next section we treat a simple model where we can perform the functional integral both exactly and in Gaussian approximation about the complex critical points and we can see how it is necessary to analytically continue the contour of functional integration to complex field configurations if one wants to correctly, perturbatively evaluate the integral.  
\section{0+1 dimensional Abelian Higgs model with Chern-Simons term\label{3}}
0+1 dimensional field theory is actually equivalent to quantum mechanics, however, we will treat this model as a field theory and continue the time $t$ to Euclidean time $t\rightarrow -i\tau$.  Here we consider $N$ massive, charged scalar fields $\phi_i$ which interact with a gauge field $A$, both of which are functions of the Euclidean time $\tau$.   The Euclidean action density is given by
\beq
{\cal L}=\sum_{i=1}^N\left(\left| (\partial_\tau +iA)\phi_i\right|^2+m^2\left|\phi_i\right|^2\right) +i\lambda A.
\eeq
where $\lambda$ is just a parameter.  We take compact Euclidean time, 
\beq
\tau\, \colon\, 0\rightarrow\beta
\eeq
which is by the usual correspondence,  the same as finite temperature $\beta =1/k_B T$.  We take the gauge choice
\beq
\partial_\tau A=0\, \Rightarrow A\equiv constant.
\eeq
Then topologically nontrivial gauge transformations $U=e^{i2\pi n\tau/\beta}$, where $n$ is an integer have the effect of shifting $A\rightarrow A-2\pi n/\beta$, which means that we can restrict $A\in 0\rightarrow 2\pi/\beta$.  The Chern-Simons term is however, not in general invariant:
\beq
i\lambda\int_0^\beta d\tau A\rightarrow i\lambda\int_0^\beta d\tau (A-\partial_\tau\Lambda )=i\lambda\int_0^\beta d\tau (A-2\pi n/\beta )=i\lambda(\beta A-2\pi n ).
\eeq
The exponential of the Chern-Simons term however, can be made invariant if the coefficient $\lambda$ is quantized to be an integer, $\lambda =N$.  We choose for  the present purpose the same integer as the number of scalar fields.  We will be considering the limit $N\rightarrow\infty$.  The critical points of the action are satisfied at the equations of motion:
\beq
-D_\tau^2\phi_i +m^2\phi_i =0\\
\int_0^\beta d\tau \sum_{i=1}^N\left(\left( i((\partial_\tau\phi_i )^*\phi_i -\phi_i^*\partial_\tau\phi_i )+2A\phi_i^*\phi_i\right)\right) -iN\beta =0.
\eeq
The second equation can easily be solved and exposes the complex critical point, the solution $A^{critical}$ is given by
\beq
A^{critical}={iN\beta -  \int_0^\beta d\tau  i\sum_{i=1}^N((\partial_\tau\phi_i )^*\phi_i -\phi_i^*\partial_\tau\phi_i )\over 2\int_0^\beta d\tau \sum_{i=1}^N\phi_i^*\phi_i}\approx {i\gamma +\alpha\over \delta}
\eeq
for some real values of $\gamma$, $\alpha$ and $\delta$.  The scalar field equation actually has no solution on the space of periodic functions for a gauge field of nontrivial holonomy, $A\ne 2n\pi /\beta$.  This 
fact is not a serious problem, since we can actually do the scalar field functional integral exactly, for any value of the gauge field, and hence eliminate it.  Expressing (for one of the $N$ scalar fields) in Fourier modes,
\beq
\phi (\tau )=\sum_{n=-\infty}^\infty\varphi_ne^{i2\pi n\tau /\beta }
\eeq
then
\beq
\int_0^\beta d\tau\left| D_\tau\phi\right|^2 =\beta\sum_{n=-\infty}^\infty\varphi_n^*\varphi_n ((2\pi n /\beta )+A)^2
\eeq
yielding each scalar field functional integral as
\beq
Z_1(\beta ,m,A)=\int_{-\infty}^\infty\prod_n d\{ \varphi_n^*\varphi_n\}e^{-\beta\sum_n\left|\varphi_n\right|^2 (({2\pi n \over\beta} )+A)^2+m^2)}.
\eeq
Integrating over the Gaussian fluctuations of each Fourier mode gives the infinite product
\beq
Z_1(\beta ,m,A)=\prod_{n=-\infty}^\infty {1\over\beta (({2\pi n \over\beta} +A)^2+m^2)} 
\eeq
which when normalized with respect to $A=0$ gives
\beq
{Z_1(\beta ,m,A)\over Z_1(\beta ,m,0)}=\prod_{n=-\infty}^\infty {({2\pi n \over\beta} )^2+m^2\over ({2\pi n \over\beta} +A)^2+m^2}.
\eeq
It is well known how to evaluate the product, adapting methods found in  \cite{j} and \cite{dll}  gives us, putting in $N$ scalar fields
\beq
{Z(\beta ,m,A)\over Z(\beta ,m,0)}=\left({\cosh\beta m -1\over \cosh\beta m -\cos\beta A}\right)^N.
\eeq
Thus the ``functional" integral that we are left with is
\beq
{\cal I}(N,\beta ,m)=\int_0^{2\pi /\beta }dA\left({\cosh\beta m -1\over \cosh\beta m -\cos\beta A}\right)^N e^{iN\beta A}.
\eeq
This integral can be performed exactly, and also in the limit $N\rightarrow\infty$ in a saddle point approximation.  The critical points of the action are achieved at complex values of the integration variable $A$.  Therefore we can compare the results that we obtain using the approximation of the saddle point obtained by considering the full complex action, and by considering only the real part of the action.  We will find that considering just the real part of the action gives wrong results.

To perform the integral exactly, we simply write the variable $z=e^{i\beta a}$.  Then $dA=dz/i\beta z$ and the integral becomes a contour integral over the unit circle in the complex $z$ plane
\beq
{\cal I}(N,\beta ,m)=\oint {dz\over i\beta}\left({2(\cosh\beta m -1)\over 2z\cosh\beta m -z^2-1}\right)^N z^N.
\eeq
The poles of the integrand are at $z_\pm=\cosh\beta m\pm\sqrt{(\cosh\beta m)^2 -1}=e^{\pm\beta m}$, hence only one of the poles contributes.  The residue is easily calculated to give the exact result:
\beq
{\cal I}(N,\beta ,m)={2\pi(\cosh\beta m -1)^N2^N\over \beta (-1)^N(N-1){\factorial}}\sum_{k=0}^{N-1}\left({N-1\atop k}\right)\left. \left({d^kz^{2N-1}\over dz^k}\right)\left({d^{N-1-k}\over dz^{N-1-k}}{1\over (z-e^{\beta m})^N}\right)\right|_{e^{-\beta m}}.
\eeq
This expression can be simplified further, but for present purposes, we will only calculate it in the limit $\beta m\rightarrow\infty$.  Then we get
\beq
{\cal I}(N,\beta ,m)\approx{e^{-N\beta m}2^{2N}\over \beta}\sqrt{\pi\over N}.\label{1}
\eeq
This result is reproduced perfectly in calculating the integral via the approximation of steepest descent.  Here the original integral is written as
\bea
{\cal I}(N,\beta ,m)&=&{(\cosh\beta m -1)^N2^N\over i\beta }\oint {dz\over z}e^{N(\ln z^2 -\ln(2z\cosh\beta m -z^2-1)}\cr
&\equiv &{(\cosh\beta m -1)^N2^N\over i\beta }\oint {dz\over z}e^{N(f(z))}
\eea
and then the method of steepest descent gives the answer:
\beq
{\cal I}(N,\beta ,m)={(\cosh\beta m -1)^N2^N\over i\beta }i{e^{N(f(z^{critical})}\over z^{critical}}\sqrt{2\pi\over N\left|f^{\prime\prime}(z^{critical})\right|}
\eeq
where $z^{critical} $ is defined by the critical point, $f^\prime (z^{critical})=0$.  Evidently, evaluating this gives exactly the answer in equation (\ref{1}), in the limit $\beta m\rightarrow\infty$.

Finally, we can also consider only the real part of the action to determine the critical point, and perform the integral in a steepest descent approximation about this abridged critical point.  The imaginary part of the action gives a bounded fluctuating contribution, which should be integrable against the measure on the space of functions, provided by the real part of the action.  It is necessary to work with the original variable $A$, and the real part of the action is just
\beq
f(A)=-\ln ( \cosh\beta m-\cos\beta A).
\eeq
This action is critical at
\beq
f^\prime (A)=-{\beta\sin\beta A\over \cosh\beta m-\cos\beta A}=0
\eeq
which implies the $A^{critical}=0,\pi/\beta, 2\pi/\beta,\cdots$.  The zero at $\pi/\beta$ is a local maximum thus does not contribute, while the zero at  $A^{critical}=0$ and at $A^{critical}=2\pi/\beta$ combine to give the full integral over the Gaussian peak,  and then using
\beq
f^{\prime\prime}(A=0)={-\beta^2\over\cosh\beta m -1}
\eeq
we get
\beq
{\cal I}(N,\beta ,m)={(\cosh\beta m -1)^N\over\beta}\sqrt{2\pi (\cosh\beta m -1)\over N}e^{-N(\cosh\beta m -1)/2}.
\eeq
It is obvious that this answer is nothing like the answer equation (\ref{1}) that we obtained by doing the proper steepest descent calculation and is evidently the wrong procedure to follow.  

\section{Conclusions\label{sec:concl}}
In conclusion, we have shown, via a specific simple example that the perturbative path integral quantization of field theories with complex critical points requires that we analytically continue into the complex extension of the space of field configurations so that the path of functional integration passes through the complex critical point according to the direction of steepest descent.  Disregarding the complex part of the action to determine the `abridged' critical points, that is the critical points of only the real part of the action, leads to erroneous results.  
\bigskip
\section{Acknowledgements\label{sec:ack}}
We thank Gerald Dunne for useful discussions and the Benasque Center for Science for very pleasant working conditions where some of this work was done.  This work was funded in part by NSERC of Canada.

\bibliographystyle{unsrt}
\bibliography{fieldtheory}

\end{document}